# Exchange anisotropy and the dynamic phase transition in thin ferromagnetic Heisenberg films

Hyunbum Jang[1], Malcolm J. Grimson[2], and Carol K. Hall[3]

[1]*Department of Physiology, Johns Hopkins University, Baltimore, MD 21205*
[2]*Department of Physics, University of Auckland, Auckland, New Zealand*
[3]*Department of Chemical Engineering, North Carolina State University, Raleigh, NC 27695-7905*

Monte Carlo simulations have been performed to investigate the dependence of the dynamic phase behavior on the bilinear exchange anisotropy of a classical Heisenberg spin system. The system under consideration is a planar thin ferromagnetic film with competing surface fields subject to a pulsed oscillatory external field. The results show that the films exhibit a single discontinuous dynamic phase transition (DPT) as a function of the anisotropy of the bilinear exchange interaction in the Hamiltonian. Furthermore there is no evidence of stochastic resonance (SR) associated with the DPT. These results are in marked contrast to the continuous DPT observed in the same system as a function of temperature and applied field strength for a fixed bilinear exchange anisotropy.

## I. INTRODUCTION

The dynamics of field-induced magnetization reversal in thin ferromagnetic films has been the subject of extensive experimental and theoretical interest [1]. For theoretical studies, the kinetic Ising model has provided a conceptually simple model to investigate the dynamic phase behavior of ferromagnets [2-5]. However, while the Ising model can provide a good representation of uniaxial ferromagnets in which magnetization reversal proceeds by nucleation and domain wall motion, it cannot account for magnetic relaxation processes such as the coherent rotation of spins. This requires a spin model with continuous degrees of freedom such as the classical Heisenberg model in which the magnetic spins can rotate through all possible orientations [6].

The dynamic phase behavior of thin ferromagnetic films of Heisenberg spins with competing surface fields subject to an applied oscillatory field was investigated in recent studies [7,8]. The inclusion of a bilinear exchange anisotropy $\Lambda$ in the model Hamiltonian allowed the system to take on Ising-like characteristics whilst allowing the magnetic spins to orient continuously. In addition, the competing surface fields ensured the presence of domains of opposite magnetization at the two film surfaces. So that the time dependence of the film magnetization in the applied oscillatory field was determined by the motion of the interface between domains of opposite magnetization. The DPT has been studied as a function of temperature, as well as the amplitude and frequency for both sinusoidal [7] and pulsed [8] applied oscillatory fields. However, these studies were limited to thin ferromagnetic films with a single value for the bilinear exchange anisotropy. This paper complements the previous studies by investigating the dependence of the dynamic phase behavior on the bilinear exchange anisotropy of the Heisenberg spin system and provides an insight into the different dynamic responses of discrete state and continuous orientation magnetic spin models.

## II. MODEL

The system under consideration here is a three dimensional thin planar film of finite thickness $D$ with competing surface fields subject to a time dependent oscillatory external field $H(t)$ with Hamiltonian

$$H(t) = H_0 - h\left(\sum_{i \in \text{surface} 1} S_i^z - \sum_{i \in \text{surface} D} S_i^z\right) - H(t)\sum_i S_i^z. \tag{1}$$

The competing surface fields are characterized by a magnitude $h$ and $H(t)$ is taken to have a pulsed form with

$$H(t) = \begin{cases} -H_0, & \dfrac{2(k-1)\pi}{\omega} < t \leq \dfrac{(2k-1)\pi}{\omega} \\ H_0, & \dfrac{(2k-1)\pi}{\omega} < t \leq \dfrac{2k\pi}{\omega} \end{cases}, \tag{2}$$

where $H_0$ is the amplitude, $\omega$ is the angular frequency and $k$ ($k = 1, 2, 3, \ldots$) is an integer representing the number of periods of the pulsed oscillatory field. The anisotropic classical Heisenberg model [9] is defined by

$$H_0 = -J\sum_{\langle i,j \rangle}\left((1-\Lambda)(S_i^x S_j^x + S_i^y S_j^y) + S_i^z S_j^z\right), \tag{3}$$

where $\mathbf{S}_i = (S_i^x, S_i^y, S_i^z)$ is a unit vector representing the $i$th spin and the notation $\langle i,j \rangle$ indicates that the sum is restricted to nearest-neighbor pairs of spins. $J > 0$ is the coupling constant for the ferromagnetic exchange interaction, while $\Lambda$ characterizes the bilinear exchange anisotropy. In the isotropic limit, $\Lambda = 0$, the model reduces to the familiar classical Heisenberg model, while for $\Lambda = 1$, the Hamiltonian becomes Ising-like.

The model film is a simple lattice of size $L \times L \times D$, in units of the lattice spacing. Periodic boundary conditions are applied in the $x$ and $y$ directions. Free boundary conditions are applied in the $z$ direction that is of finite thickness $D$. The system is subject to competing applied surface fields of magnitude $h = -0.55$ in layers $n = 1$ and $n = D$ of the film. A film thickness $D = 12$ was used throughout. This value corresponds to the crossover regime between wall and bulk dominated behavior [10] for which the equilibrium phase behavior of the system is well characterized [11,12]. The results reported here are for lattices of size $L = 32$.

Monte Carlo simulations were performed using the Metropolis algorithm [13] with a random spin update scheme. Trial configurations were generated by the rotation of a randomly selected spin through a random angular displacement about one of the $x$, $y$, $z$ axes chosen at random [14,15]. A sequence of size $L \times L \times D$ trials comprises one Monte Carlo step per spin (MCSS), the unit of time in our simulations. The period of the pulsed oscillatory external field is given by product $R_{FS} \times N$, where $R_{FS}$ is the field sweep rate and $N$ is a number of MCSS. The applied oscillatory field $H(t)$ being updated after every MCSS according to Eq. (2). The simulations reported here were performed for a value of $R_{FS} = 1$ with $N = 240$. In all of the simulations the initial spin configuration was a ferromagnetically ordered state of the spins with $S_i = +1$ for all $i$.

The order parameter for the DPT is the period averaged magnetization over a complete cycle of the pulsed field, $Q$, defined by

$$Q = \frac{\omega}{2\pi} \oint M_z(t)\, dt . \qquad (4)$$

where the $z$-component of the magnetization for the film is

$$M_z(t) = \frac{1}{D} \sum_{n=1}^{D} M_n^z(t) , \qquad (5)$$

with

$$M_n^z(t) = \frac{1}{L^2} \sum S_i^z(t) \qquad (6)$$

being the $z$-component of the magnetization for the $n$th layer of the film. The system exhibits a dynamically ordered phase with $|Q| > 0$ and a dynamically disordered phase with $Q = 0$. The period averaged magnetization for the $n$th layer of the film is given by

$$Q_n = \frac{\omega}{2\pi} \oint M_n^z(t)\, dt . \qquad (7)$$

### III. RESULTS

The mean period averaged magnetization, $\langle Q \rangle$, as a function of the bilinear exchange anisotropy, $\Lambda$, is shown in Fig. 1 for two different sets of the external field amplitude $H_0$ and reduced temperature $T^* = k_B T/J$: $H_0 = 1.0$ with $T^* = 0.6$ (open symbols) and $H_0 = 0.55$ with $T^* = 1.0$ (solid symbols). The quantity $\langle Q \rangle$ is determined from a sequence of full cycles with initial transients discarded. The error bars in the figure correspond to a standard deviation in the measured values and are only visible when they exceed the size of the symbol. The lines in the figure are only to guide the eye. The DPT is characterized by the vanishing of the order parameter $Q$ at a value of $\Lambda$. For $H_0 = 1.0$ and $T^* = 0.6$, $\langle Q \rangle$ vanishes at a value of $\Lambda = 0.18$, while for $H_0 = 0.55$ and $T^* = 1.0$, $\langle Q \rangle$ vanishes at a value of $\Lambda = 0.30$. However the most remarkable feature of Fig. 1 is that the film shows a discontinuous DPT as a function of $\Lambda$. Although it should be noted that the fluctuations in $\langle Q \rangle$ close to the DPT are very large as indicated by the size of the error bar. This is in marked contrast to the dynamic phase behavior of these films as a function of both $T^*$ and $H_0$ for a fixed $\Lambda$ where the DPT was continuous [8].

Fluctuations of the order parameter $\chi(Q)$ were measured in the simulations with

$$\chi(Q) = L^2 D \left( \langle Q^2 \rangle - \langle |Q| \rangle^2 \right), \tag{8}$$

where $\langle \ \rangle$ denotes the average over a sequence of full cycles with initial transients discarded, and $L^2 D$ is the number of spins in the system. Note that the absolute order parameter, $|Q|$, is used in the definition of $\chi(Q)$ since in the dynamically ordered phase the probability density for $Q$ has peaks at both $+Q$ and $-Q$ [4]. Following Kim *et al.* [16] evidence for stochastic resonance (SR) at the DPT is obtained from measurement of the occupancy ratio $Q^{OR}$ defined by

$$Q^{OR} = \frac{\omega}{2\pi} \oint M_z(t) \frac{H(t)}{|H(t)|} \, dt, \tag{9}$$

where $H(t)/|H(t)|$ is the sign of the external pulsed oscillatory field.

Fig. 2. shows the fluctuations in the dynamic order parameter, $\chi(Q)$, and the mean period averaged occupancy ratio, $\langle Q^{OR} \rangle$, as a function of $\Lambda$ for $H_0 = 1.0$ at $T^* = 0.6$ (open symbols) and for $H_0 = 0.55$ at $T^* = 1.0$ (solid symbols). The fluctuations in $Q$ show a single peak centered on a value for $\Lambda$ corresponding to the discontinuity in $\langle Q \rangle$ seen in Fig. 1. However, of more interest is the featureless form for $\langle Q^{OR} \rangle$ at values of $\Lambda$ corresponding the peak in $\chi(Q)$, indicating that stochastic resonance (SR) is not associated with the discontinuous DPT seen as a function of $\Lambda$. Taken together with the results of a previous study [8], this suggests that the DPT observed in thin ferromagnetic films with competing surface fields is not related to any occurrence of SR. This is noteworthy, since for the corresponding free film and bulk system, the DPT is seen to be associated with SR.

More detail on the nature of the DPT in the thin film with competing surface is available from the form of the dynamic order parameter in the layers of spins across the film. Figure 3 shows the bilinear exchange anisotropy dependence of the dynamic order parameter for the *n*th layer, $Q_n$, across the whole film for (a) $H_0 = 1.0$ with $T^* = 0.6$, and (b) $H_0 = 0.55$ with $T^* = 1.0$. For large values of $\Lambda$ ($\Lambda > 0.2$ in Fig. 3(a) and $\Lambda > 0.35$ in Fig. 3(b)) the film is in a dynamically ordered phase with $\langle Q \rangle > 0$. This is a result of the layer dynamic order parameter being non-zero and essentially uniform across the film except close to one surface. As $\Lambda$ decreases to a critical value ($\Lambda = 0.18$ in Fig. 3(a) and $\Lambda > 0.30$ in Fig. 3(b)) there is an abrupt change in $\langle Q_n \rangle$ across the whole film that is particularly marked in the bulk of the film. Notably, this is abrupt change in $\langle Q_n \rangle$ located at the same value of Λ for each layer of the film, a value that is equivalent to the transition value of Λ for the DPT in the whole film obtained from Fig. 1. For small values of Λ (Λ < 0.18 in Fig. 3(a) and Λ < 0.30 in Fig. 3(b)) the layer dynamic order parameter across the film is antisymmetric about the mid-point of the film and corresponds to a dynamically disordered phase for the whole film with $\langle Q \rangle = 0$. The results of Fig. 3 clearly show that there is a single DPT as a function of Λ for the film with the surface layers of spins undergoing a DPT at the same Λ value as the bulk spins. This is in contrast to the results for the film with competing surface fields with a fixed Λ, where the DPT for the surface layers of spins differed from the DPT for the spins in the bulk of the film as a function of both $T^*$ and $H_0$ [8].

## IV. DISCUSSION

The dynamic response of a ferromagnet to an oscillatory external field can be viewed as a competition between two time scales: the half-period of the external field that is proportional to the inverse driving frequency and the average metastable lifetime of the system after a sudden field reversal. At low driving frequencies the time dependent magnetization oscillates about zero with the external field (symmetric dynamic phase). For high frequencies, however, the magnetization does not have time to switch sign during one half period of the external field and so oscillates about one or the other of its degenerate zero-field values (asymmetric dynamic phase). This symmetry breaking corresponds to a DPT and numerous studies of the kinetic Ising model [2-5] have shown a continuous DPT between the symmetric and asymmetric dynamic phases.

Korniss *et al* [5] have demonstrated that the metastable lifetime of the kinetic Ising model after a sudden field reversal depends on the temperature and the field amplitude. For a sufficiently large system, the kinetic Ising model escapes from the metastable phase through the nucleation of many droplets and subsequently the time-dependent system magnetization is self-averaging. But for any finite system the metastable decay mode changes to the nucleation and growth of a single droplet at sufficiently low temperatures. Due to the stochastic nature of the nucleation of a single droplet, the corresponding response of the system in the presence of an oscillatory field is different and the system exhibits SR. For an infinitely large system a continuous DPT should persist down to an arbitrarily low temperature. But, in a finite system the DPT gives way to SR that can be misinterpreted as indicating the existence of a discontinuous DPT [17].

Dynamic Monte Carlo studies of the anisotropic Heisenberg model [7,8] show a continuous DPT as a function of temperature and field amplitude for both the bulk system and the free film. Furthermore, the DPT is associated with SR as suggested by Korniss *et al*. However in the thin film with competing surface fields there is no evidence of SR associated with the continuous DPT. This is a result of the static competing surface fields that ensure the system is always in a "single droplet" regime, since in all but very strong oscillatory fields the two phases always coexist within the film. The DPT thus proceeds simply by the growth of one phase through domain wall motion as a result of coherent spin rotation of the Heisenberg spins.

The results in this paper for the anisotropic Heisenberg model in thin films with competing surface fields show a discontinuous DPT. However, the discontinuous DPT only occurs as function of the anisotropy of the bilinear exchange interaction in the Hamiltonian. The observed discontinuous DPT is thus related to the crossover in the dynamic response of the model from that of an Ising-like spin system to that of a classical Heisenberg spin system. For a fixed value of $\Lambda$, a continuous DPT as a function of temperature, field amplitude and frequency is found.

**Acknowledgments**

The authors would like to thank Thomas B. Woolf for his support and interest in this work.

**Figure Captions**

**Fig. 1** Mean period averaged magnetization, $\langle Q \rangle$, as a function of the bilinear exchange anisotropy, $\Lambda$, for $H_0 = 1.0$ and $T^* = 0.6$ (open symbols) and for $H_0 = 0.55$ and $T^* = 1.0$ (solid symbols).

**Fig. 2** (a) Fluctuations of the dynamic order parameter, $\chi(Q)$, and (b) the mean period averaged occupancy ratio, $\langle Q^{OR} \rangle$, as a function of the bilinear exchange anisotropy, $\Lambda$. Open symbols correspond to $H_0 = 1.0$ and $T^* = 0.6$, while solid symbols represent $H_0 = 0.55$ and $T^* = 1.0$.

**Fig. 3** Period averaged magnetizations for the $n$th layer, $Q_n$, across the whole film as a function of the bilinear exchange anisotropy, $\Lambda$, for (a) $H_0 = 1.0$ and $T^* = 0.6$ and (b) $H_0 = 0.55$ and $T^* = 1.0$.

Fig. 1

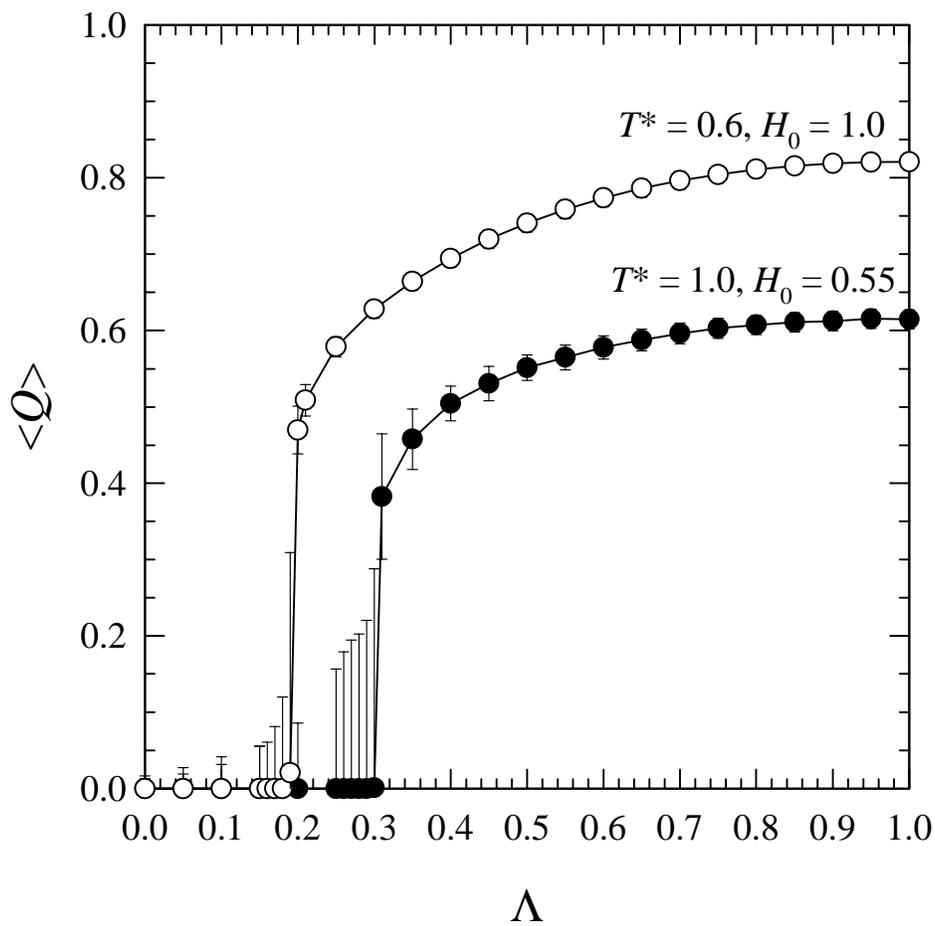

Fig. 2(a)(b)

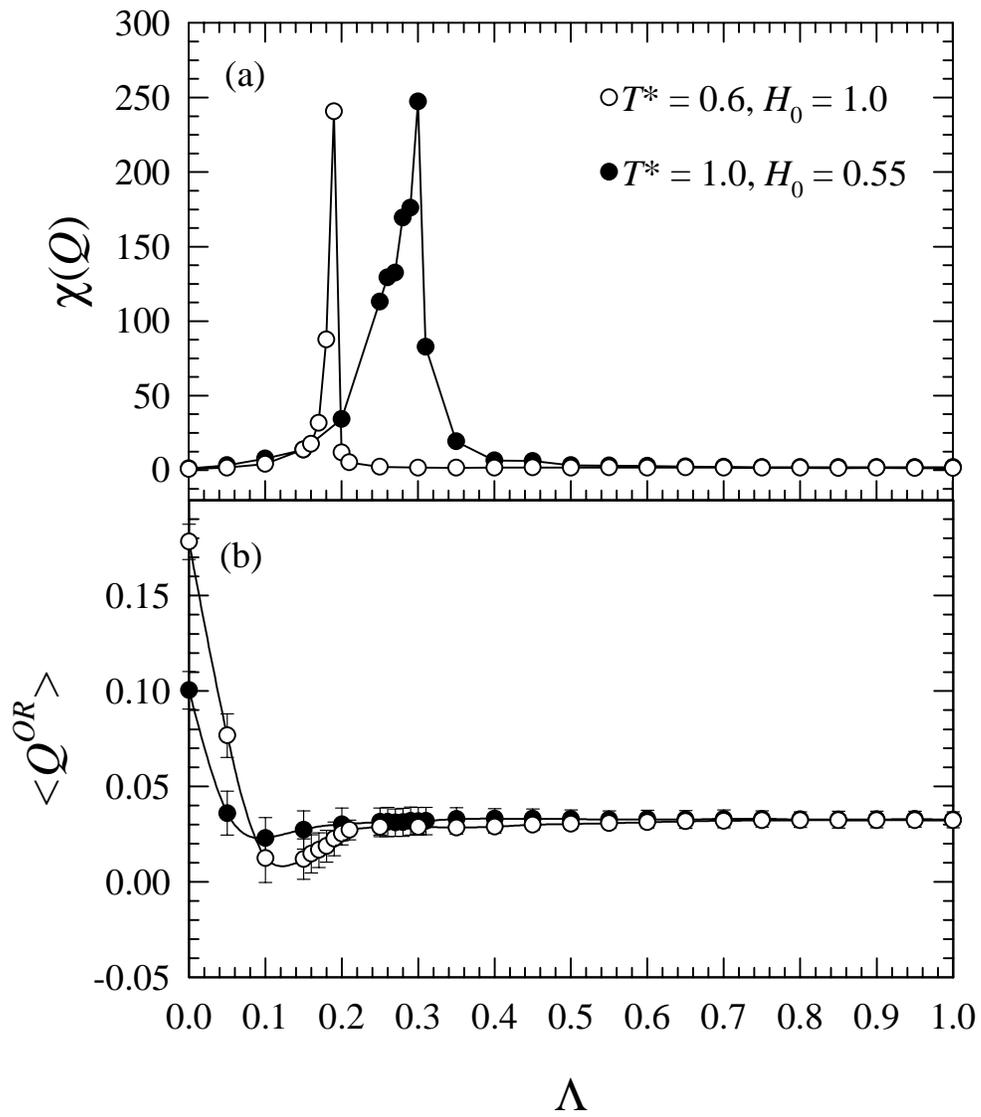

Fig. 3(a)

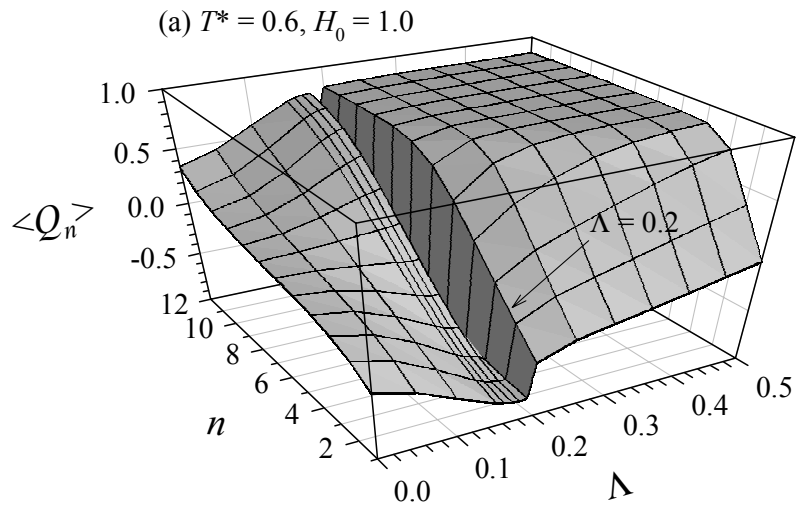

Fig. 3(b)

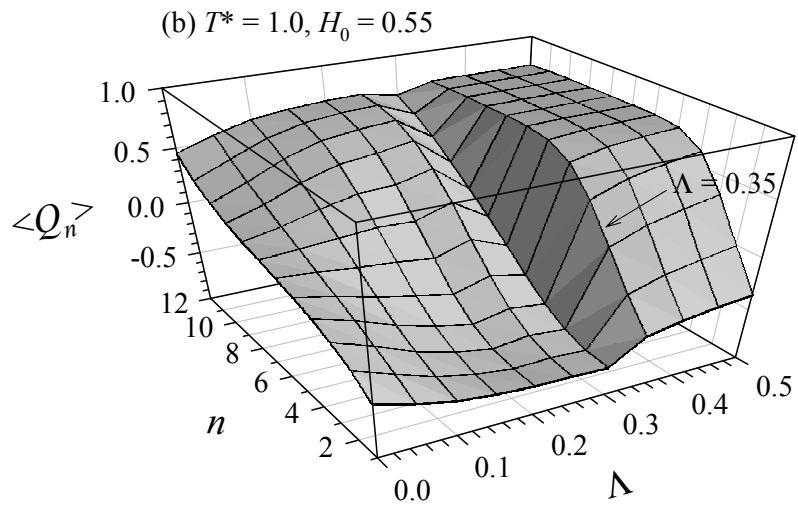